\documentclass[11pt,preprint]{aastex}

\usepackage{natbib}
\newcommand{\bd}{BD+44$^{\circ}$493}
\newcommand{\kms}{~km~s$^{-1}$} 
\newcommand{\teff}{$T_{\rm eff}$}
\newcommand{\logg}{$\log g$}

\newcommand{\vt}{$v_{\rm micro}$}

\slugcomment{Molecular features of {\bd}}

\shorttitle{Molecular features of {\bd}}
\shortauthors{Aoki et al.}

\begin{document}

\title{Molecular line formation in the extremely metal-poor star {\bd}\altaffilmark{1}}



\author{Wako Aoki \altaffilmark{2,3}}
\altaffiltext{1}{Based on data collected at the Subaru Telescope, which is operated by the National Astronomical Observatory of Japan}
\altaffiltext{2}{National Astronomical Observatory of Japan, Mitaka, Tokyo,
181-8588 Japan; aoki.wako@nao.ac.jp}
\altaffiltext{3}{Department of Astronomical Science, The Graduate
  University of Advanced Studies, Mitaka, Tokyo, 181-8588 Japan}




\begin{abstract}

Molecular absorption lines of OH (99 lines) and CH (105 lines) are
measured for the carbon-enhanced metal-poor star {\bd} with
[Fe/H]$=-3.8$. The abundances of oxygen and carbon determined from
individual lines based on an 1D-LTE analysis exhibit significant
dependence on excitation potentials of the lines; $d\log
\epsilon/d\chi \sim -0.15$--$-0.2$ dex/eV, where $\epsilon$ and $\chi$
are elemental abundances from individual spectral lines and their
excitation potentials, respectively. The dependence is not explained
by the uncertainties of stellar parameters, but suggests that the
atmosphere of this object possesses a cool layer that is not
reproduced by the 1D model atmosphere. This result agrees with the
predictions by 3D model calculations. Although absorption lines of
neutral iron exhibit similar trend, it is much weaker than found in 
molecular lines and that predicted by 3D LTE models.  

\end{abstract}





\section{Introduction}

Structures of stellar atmospheres have been traditionally modeled as a
function of depth assuming static and horizontally homogeneous 
stratification \citep[e.g., ][]{gustafsson75, kurucz79}. Real stellar
atmospheres should, however, have three-dimensional (3D) hydrodynamic
structures.  Recent efforts for developing 3D hydrodynamic models of
stellar atmospheres have been predicting the effects
on the wavelength dependence of the limb darkening and wavelength
shift of absorption features that are able to be examined for the Sun
\citep[e.g., ][]{asplund09}.

The 3D effects on the structure of stellar atmospheres are expected to
be much larger in metal-poor stars than objects with the solar
metallicity. A large impact is expected on the surface layers, in
which the temperature predicted by the 3D models is significantly
lower than that by 1D models. This is due to the strong adiabatic
cooling in 3D models, which is mostly compensated by the line
blanketing by numerous spectral lines in metal-rich stars
\citep{asplund99}. The average temperature of the surface layer of 3D
models for stars with [Fe/H]$\sim -3$ is predicted to be about 1000~K
lower than than that of corresponding 1D models \footnote{[A/B] =
  $\log(N_{\rm A}/N_{\rm B}) -\log(N_{\rm A}/N_{\rm B})_{\odot}$, and
  $\log\epsilon_{\rm A} =\log(N_{\rm A}/N_{\rm H})+12$ for elements A
  and B.}. Since metal-poor stars are not spatially resolved in
general, and the absolute radial velocity is not determined as for the
Sun, the 3D effects cannot be examined by the limb darkening nor the
line shift.

The effect of the cool surface layer is, however, expected to appear
in absorption features in atomic and molecular lines. Since molecule
formation is promoted in cool layers, molecular lines become stronger
in general in the calculations using 3D model atmospheres. In other
words, chemical abundances derived from molecular lines (e.g., the carbon
abundance derived from CH molecular lines) using 3D model atmosphere
is significantly lower than that derived using 1D models. A similar trend is
expected for Fe lines. Whereas the ionized Fe lines are not sensitive
to the 3D effects, the Fe abundance derived from its neutral species
using an 1D model could be significantly overestimated
\citep{asplund05}.

Moreover, the effect is dependent on the excitation
potential \footnote{Excitation potential means lower atomic or
  molecular energy level of a transition in this paper, although the
  term 'lower excitation potential' is sometimes used for the same
  meaning. In order to avoid confusion with the term `lower
  excitation potential', we express it smaller, rather than lower,
  when the value of the excitation potential is smaller.} of
absorption lines, i.e., lines with smaller excitation potential are
formed in cool surface layers that appear in 3D models. If atmospheres
of metal-poor stars have cool layers in reality as predicted by 3D
models, a significant dependence of the derived abundances from
individual lines on excitation potential is expected in the
results obtained by 1D models. This can be an excellent examination for the
3D effects \citep{asplund03}. Quantitative estimates of this effect
from 3D models were made by \citet{collet07} for Fe lines and by
\citet{frebel08} for CH, NH, and OH molecular features. More recently,
\citet{dobrovolskas13} also investigated line formation in atmospheres
of giants close to the base of the red giant branch using their 3D
models.

\citet{frebel08} estimated the effect for molecular features
of HE~1327$-$2326, the most iron-deficient star known at that
time. Although some signature of the 3D effects is suggested for the
CH and NH absorption features, that is not very clear because
only molecular bands are observed in this object in which spectral
lines with different excision potentials overlap each other.

The extremely metal-poor (EMP) subgiant {\bd} provides an excellent
data-set of absorption lines that is useful to examine the 3D
effects. This star was reported by \citet{ito09} as a carbon-enhanced
object with extremely low metallicity ([Fe/H]$=-3.8$). Detailed
analysis of the spectrum of this object was recently reported by
\citet{ito13} in which a spectral atlas is provided. The spectrum
exhibits a number of clean CH and OH molecular lines. The apparent
brightness of this object (the $V$ magnitude of 9.1) and appropriate
effective temperature ({\teff} of 5430~K) make it possible to measure
accurately individual molecular lines even for OH lines that appear
around 3100~{\AA}.

In this paper, we measure OH and CH molecular lines of {\bd} and
determine the oxygen and carbon abundances from individual spectral lines using
1D, static model atmospheres. We investigate the dependence of the
derived abundances on excitation potentials of spectral lines as a
test of 3D effects on this EMP star.

\section{Observational data and line lists}

The high resolution spectrum of {\bd} analyzed in the present work was
obtained with the High Dispersion Specrograph \citep[HDS;
][]{noguchi02} of the Subaru Telescope. The data were studied by
\citet{ito09, ito13} and the details on the observation are reported
in these papers. The spectral resolution is $R=90,000$ that is
sufficiently high for measurements of molecular lines. The
signal-to-noise ratios (per pixel) around 3100~{\AA} and 4000~{\AA},
in which OH and CH molecular lines exist, are 100 and 400,
respectively.

\begin{figure}
\epsscale{.70}
\plotone{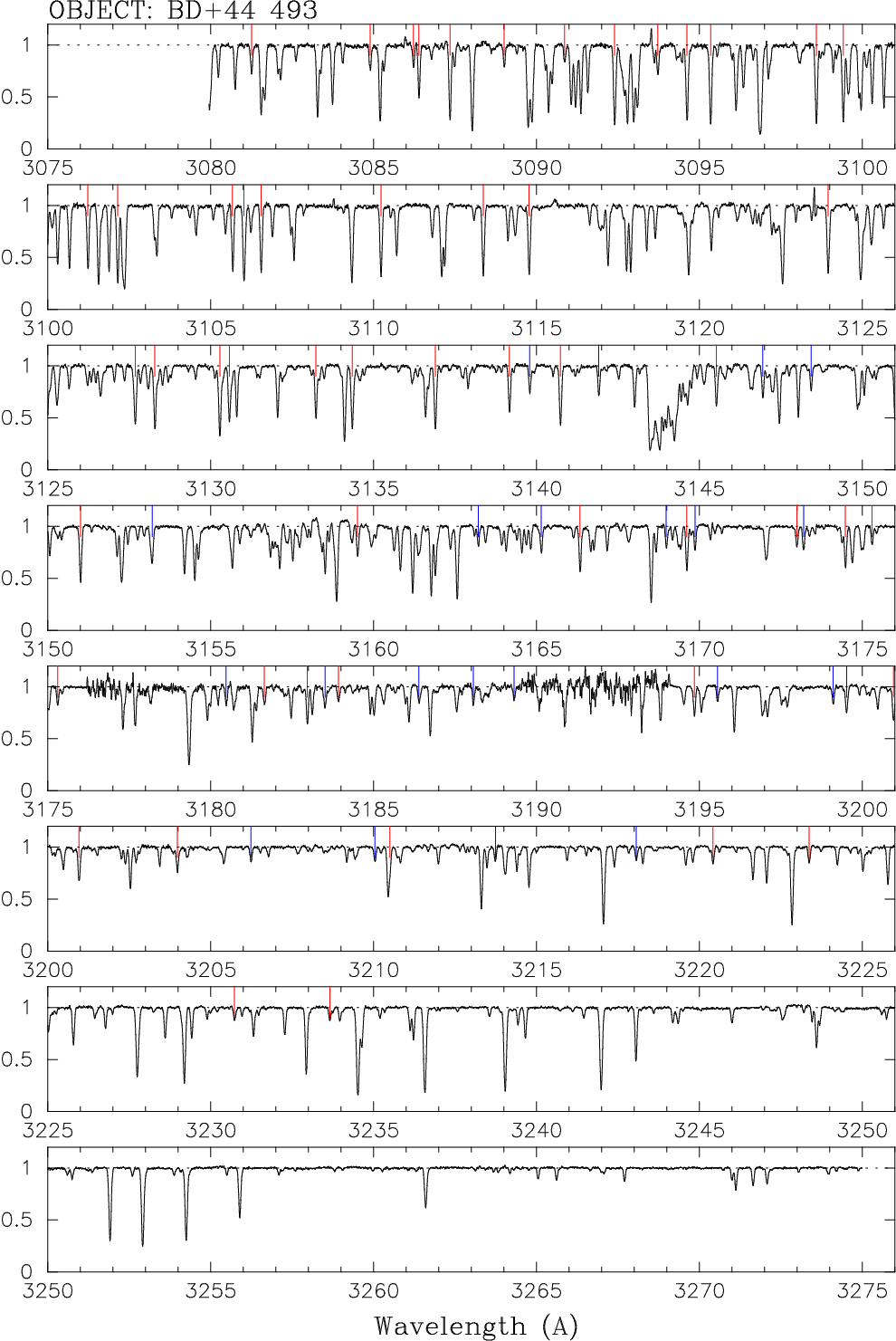}
\caption{The spectrum of {\bd} in the near UV region. The vertial
  lines indicate the positions of spectral lines of the OH $A-X$
  systems used in the analysis (red: the 0--0 system; blue: the 1--1
  system; black: the 2--2 system).
\label{fig:spoh}}
\end{figure}

\begin{figure}
\epsscale{.70}
\plotone{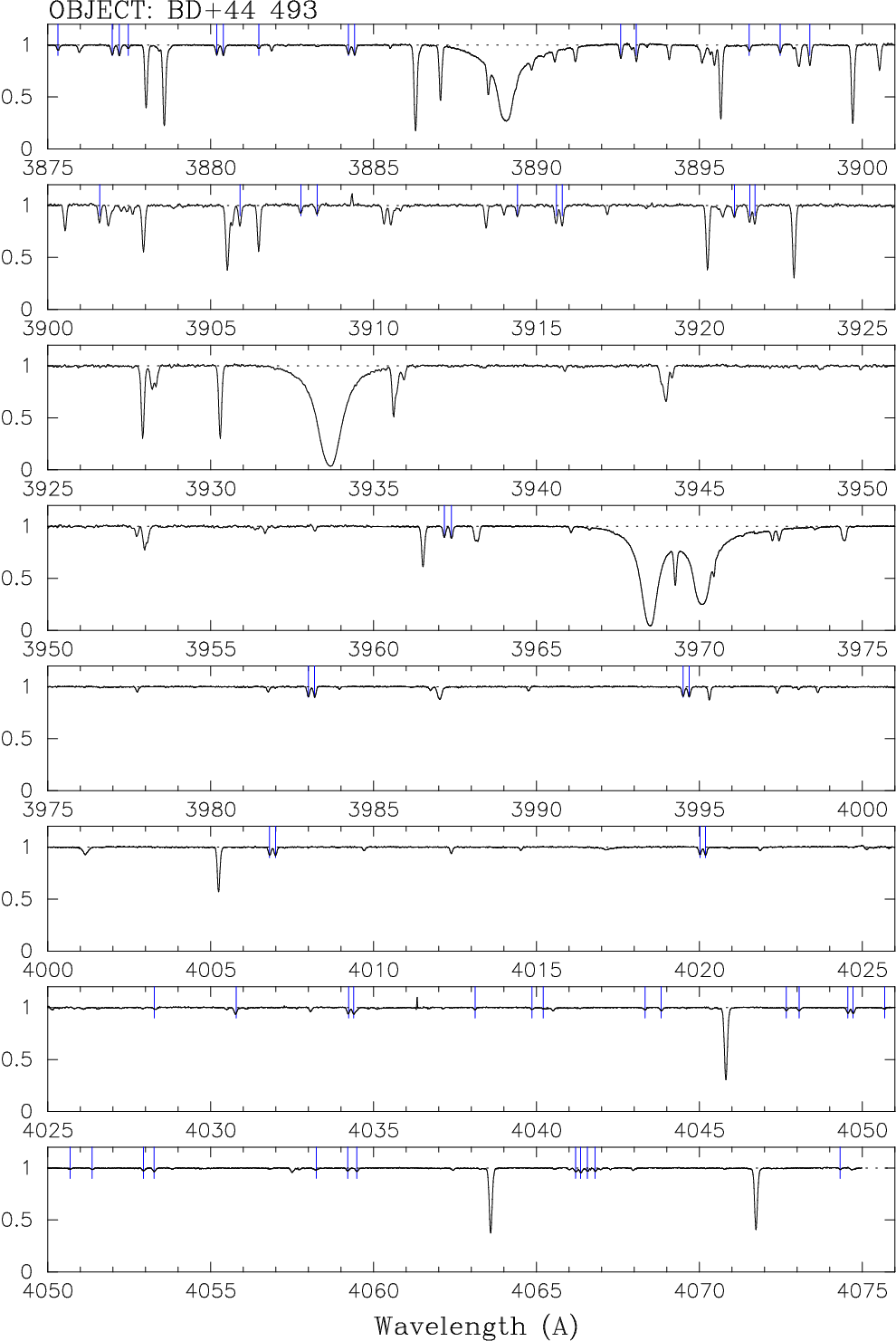}
\caption{The same as Figure~\ref{fig:spoh}, but for the CH $B-X$
  system. 
 \label{fig:spch}}
\end{figure}

Examples of spectra of OH $A-X$ and CH $B-X$ systems are shown in
Figures~\ref{fig:spoh} and \ref{fig:spch}, respectively. We selected
99 OH lines and 105 CH lines for measurements of equivalent widths by
Gaussian fitting. The line list of OH and CH, along with measured
equivalent widths are given in Tables~\ref{tab:oh} and \ref{tab:ch},
respectively.


The line positions and excitation potentials of the OH $A-X$ 0--0,
1--1 and 2--2 systems are adopted from the line list compiled by
Kurucz\footnote{http://kurucz.harvard.edu/molecules.html}. The
oscillator strengths ($gf$-values) in Table~\ref{tab:oh} are derived
from the transition probabilities ($A$-coefficient) of \citet{luque98}
that are taken from
LIFBASE\footnote{http://www.sri.com/engage/products-solutions/lifbase}. The
source of the transition probability is adopted in the compilation by
\citet{gillis01} that has been used by previous studies on oxygen abundances
of metal-poor stars based on OH $A-X$ lines \citep[e.g.,
][]{garciaperez06, frebel08}.



The line positions, excitation potentials, and transition
probabilities of the CH $B-X$ 0--0 and 1--1 bands are taken from
\citet{masseron14}, who adopted the lifetimes measured by
\citet{luque96chax,luque96chbx} to obtain the $gf$ values. The
$gf$-values are higher by 0.15~dex on average than those of
\citet{aoki02} who adopted the $f$ values of the CH $B-X$ systems from
\citet{brown87}.


\section{Analysis and results}\label{sec:ana}

\subsection{Abundance analysis based on 1D/LTE model atmospheres}

We apply model atmospheres of the ATLAS NEWODF grid \citep{castelli97}
to our 1D/LTE analysis as done by \citet{ito13}. Atmospheric
parameters derived by \citet{ito13} are adopted: {\teff}$=5430$~K,
{\logg}$=3.4$, [Fe/H]$=-3.8$, and {\vt}$=1.3~${\kms}.  We also examine
the effect of the changes of {\teff} and {\vt} on the results below.

\subsection{OH lines}

\begin{figure}
\epsscale{.70}
\plotone{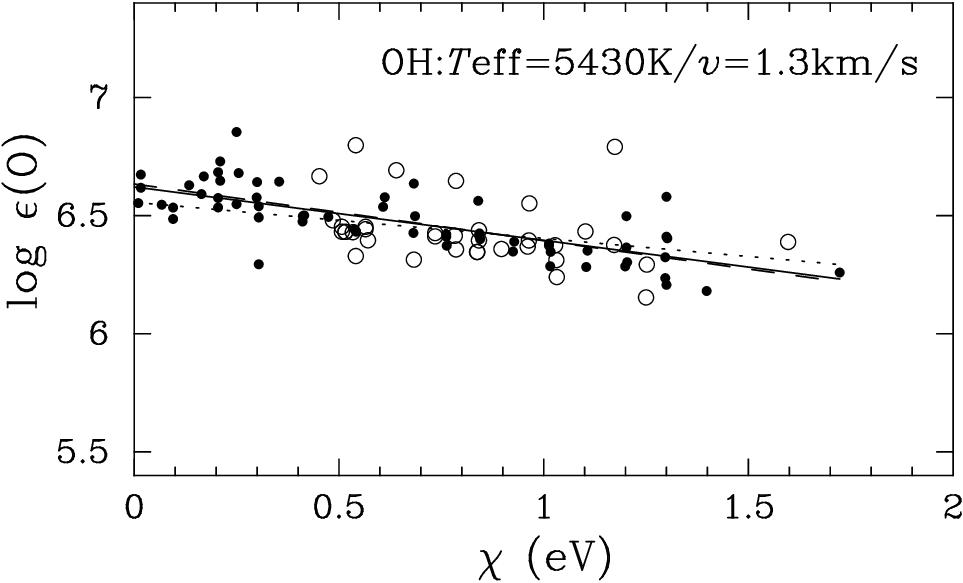}
\caption{Oxygen abundances derived from individual OH lines as a function of their excitation potential ($\chi$). The results for lines of the OH $A-X$ 0--0 and 1--1 systems are presented by filled and open circles, respectively. The results of linear fit for the two systems are shown by dashed and dotted lines, respectively. The solid line indicates the fit for all data points. \label{fig:oh3}}
\end{figure}

Figure~\ref{fig:oh3} depicts the O abundances derived from individual OH
lines as a function of their excitation potential. The results from
different systems are shown by different symbols. The derived
abundances exhibit a clear dependence on the excitation potential. A
least square fit to the data points indicates the slope of
$d\log\epsilon/d\chi = -0.23 \pm 0.03$ dex/eV. A similar result is
obtained even if the analysis is limited to the lines of the $A-X$
0--0 system: the slope is $-0.24 \pm 0.03$ dex/eV. The null hypothesis
that there is no correlation between the two values is rejected by the
regression analysis at the 99.9\% confidence level. The slope derived
from the 1--1 system is $-0.15 \pm 0.08$ dex/eV. Although the
correlation derived from the 1--1 system alone is slightly weaker, the
confidence level is still 95\% by the same analysis.

%
%

\begin{figure}
\epsscale{.40}
\plotone{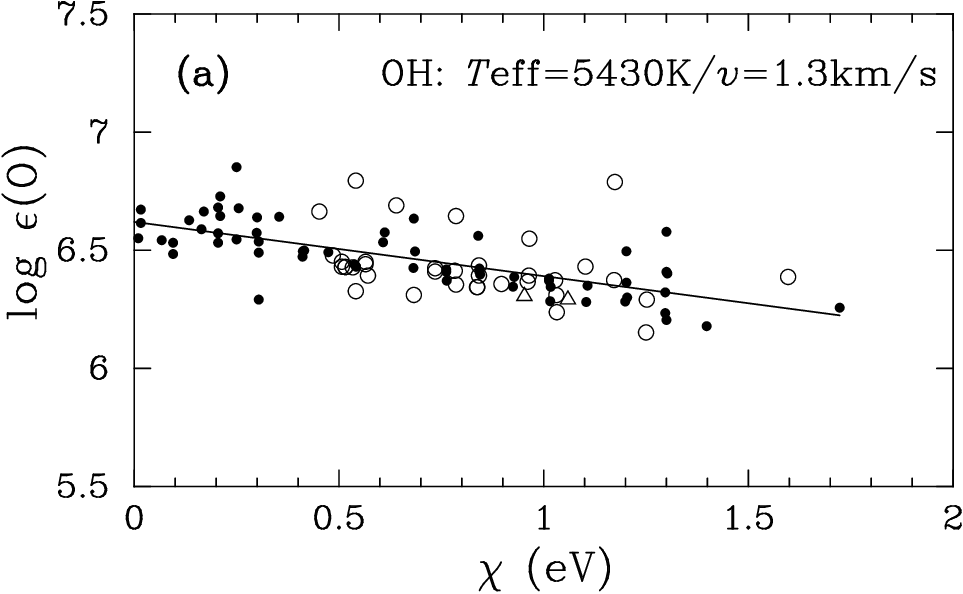} 
\plotone{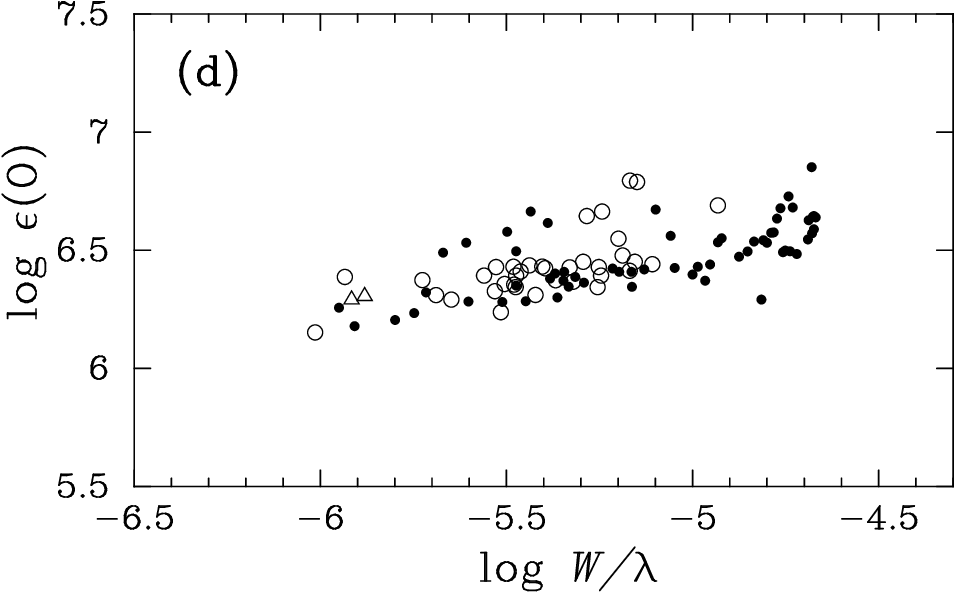} 
\plotone{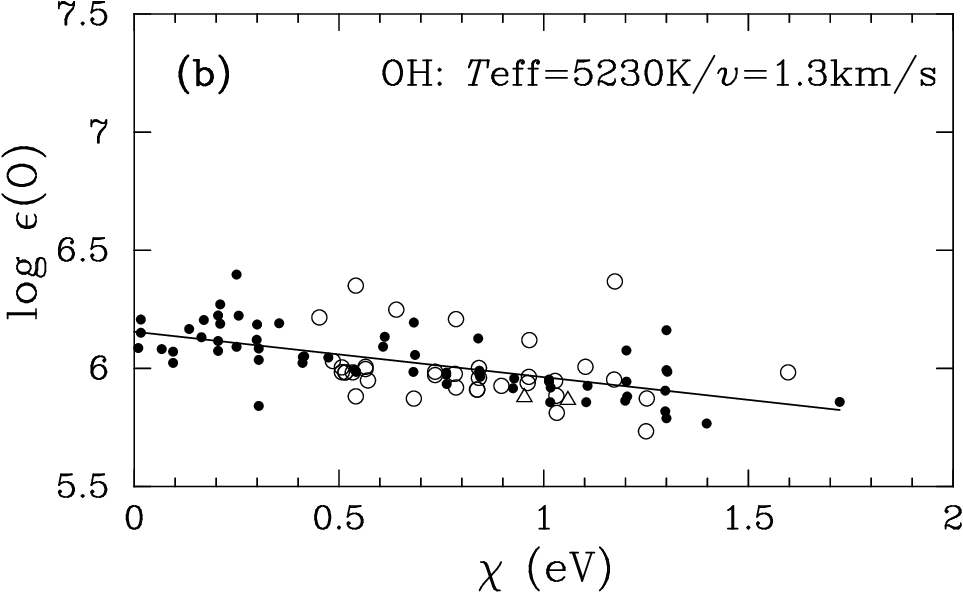} 
\plotone{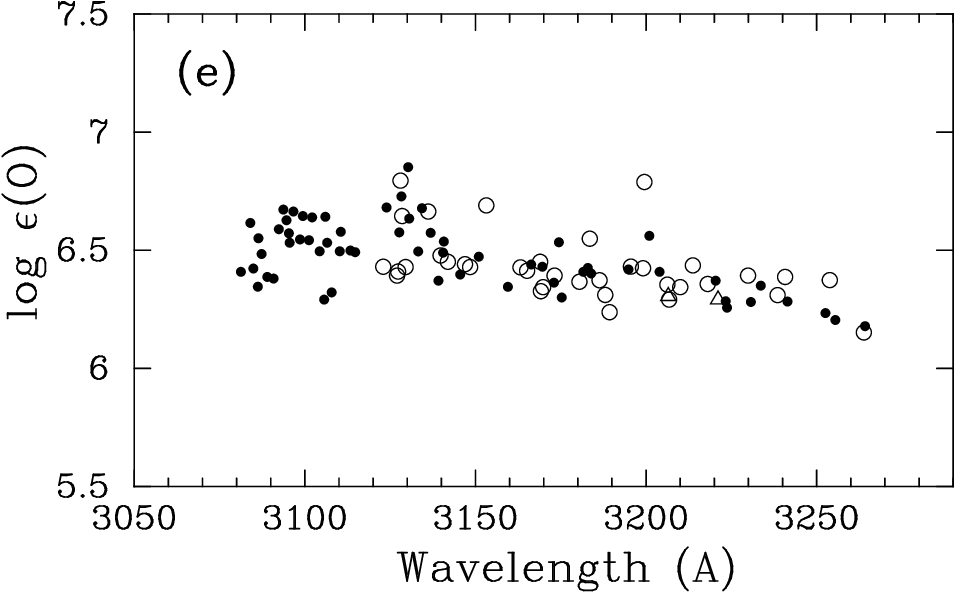} 
\plotone{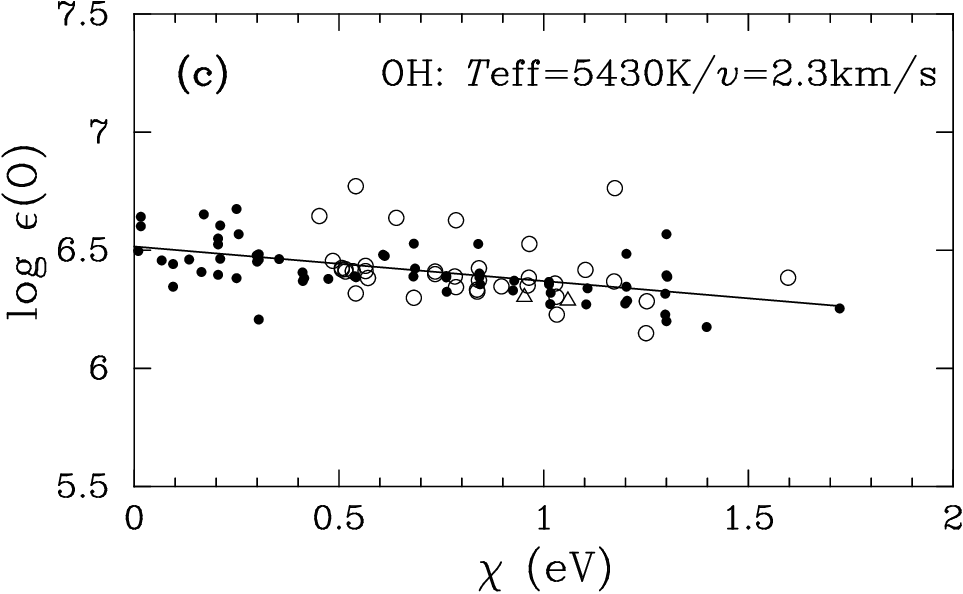} 
\plotone{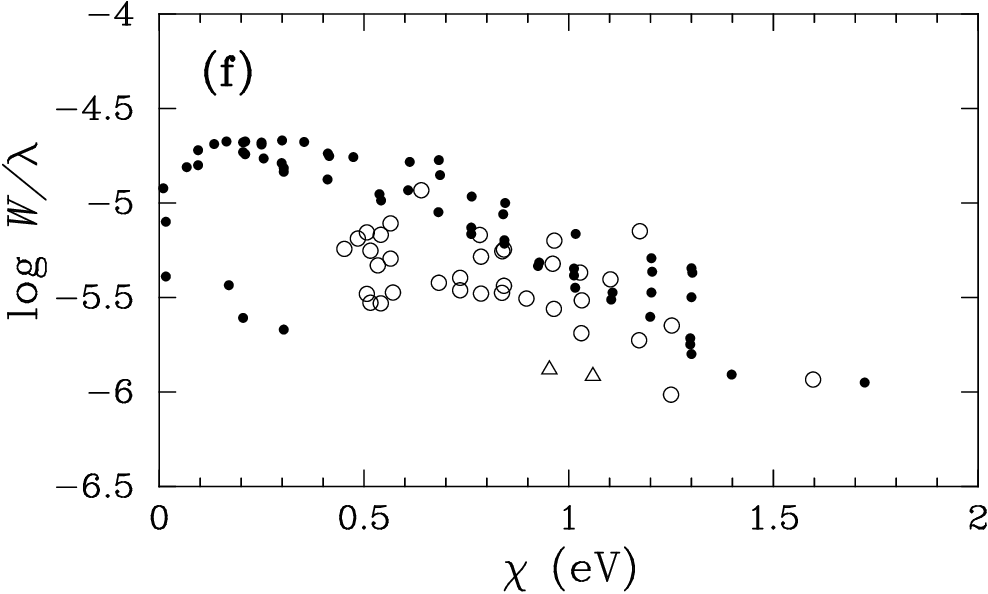} 
\caption{Analysis results for OH lines (filled circles: $A-X$ 0--0
  system; open circles: $A-X$ 1--1 system; open triangles: $A-X$ 2--2
  system). (a-c) Derived oxygen abundances, as a function of the
  excitation potentials of OH lines, are shown for effective
  temperature and micro-turbulent velocity presented in each
  panel. The trend is found even if a lower effective temperature or
  larger micro-turbulent velocity is assumed. Derived oxygen
  abundances are shown as a function of reduced equivalent widths
  ($W/\lambda$) (d) and wavelengths (e) of OH lines. Weak trends are
  found in these plots, but they can be attributed to the correlation
  between the excitaion potentails and equivalent widths shown in
  (f) and that between excitation potentials and wavelenths. See text
  for more details. \label{fig:oh}}
\end{figure}

The correlation between the derived oxygen abundances and the
excitation potentials is dependent on the adopted stellar parameters,
in particular the effective temperature and the micro-turbulent
velocity. If a lower effective temperature is adopted, the spectral
lines with larger excitation potentials become weaker, resulting in
higher abundances derived from these lines to explain the observed
line absorption. Indeed, the slope in the diagram of the oxygen
abundance and excitation potential is $-0.16 \pm 0.03$ dex/eV for the
analysis assuming {\teff}$=5230$~K (Fig. \ref{fig:oh}b). Although the
correlation is weaker than the that derived for the effective
temperature we adopt ({\teff}$=5430$~K), it is still statistically
significant. We note that the oxygen abundance derived for
{\teff}$=5230$~K on average is about 0.4~dex lower than that for
{\teff}$=5430$~K.



The slope is also dependent on the assumed micro-turbulent
velocity. The micro-turbulent velocity is determined as the elemental
abundances derived from individual lines show no dependence on the
line strengths. The micro-turbulent velocity of this object,
{\vt}$=1.3$ {\kms}, is determined by the analysis of \ion{Fe}{1} lines
\citep{ito13}. Adopting this value, a correlation between the oxygen
abundances and the strengths of OH lines is found
(Fig.~\ref{fig:oh}d). This is, however, attributed to the fact that
lines with smaller excitation potential are stronger in general
(Fig.~\ref{fig:oh}f). There are four exceptionally weak lines ($\log
W/\lambda \sim -5.5$) among those with small excitation potential
($\chi \lesssim 0.3$ eV). The oxygen abundances derived from these
four lines are as high as the value obtained from stronger lines
($\log \epsilon$(O) $\sim 6.5$), supporting our interpretation that
the apparent correlation between the oxygen abundances and the
strengths of OH lines is due to the dependence of the derived
abundances on the excitation potentials of the OH lines. The slope
$d\log\epsilon/d\chi$ obtained for {\vt}$=1.3$ {\kms} and
{\teff}$=5430$~K is $-0.12 \pm 0.03$ (Fig. \ref{fig:oh}c).


We note for completeness that, although a weak correlation is found
between the derived abundances and wavelengths of the lines
(Fig.~\ref{fig:oh}e), it is also explained by the fact that most of
the lines with smaller excitation potential, from which higher abundances are derived,
exist at shorter wavelengths.

\subsection{CH lines}

\begin{figure}
\epsscale{.40}
\plotone{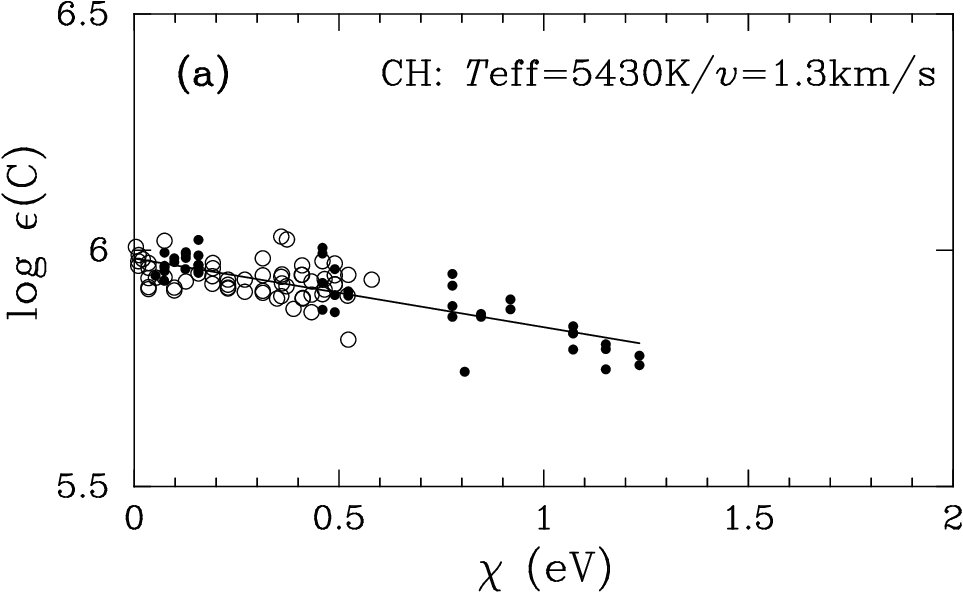} 
\plotone{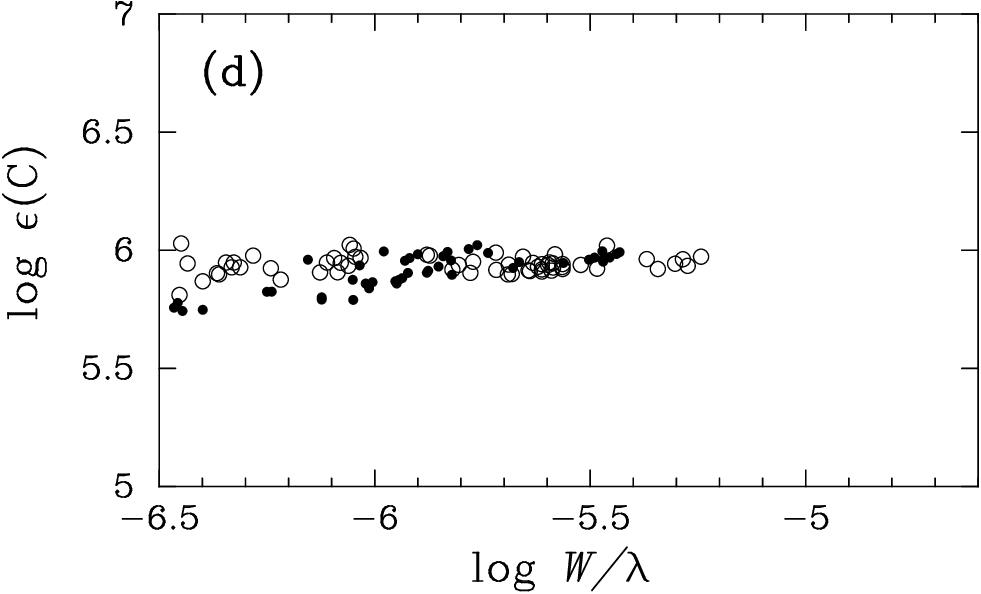} 
\plotone{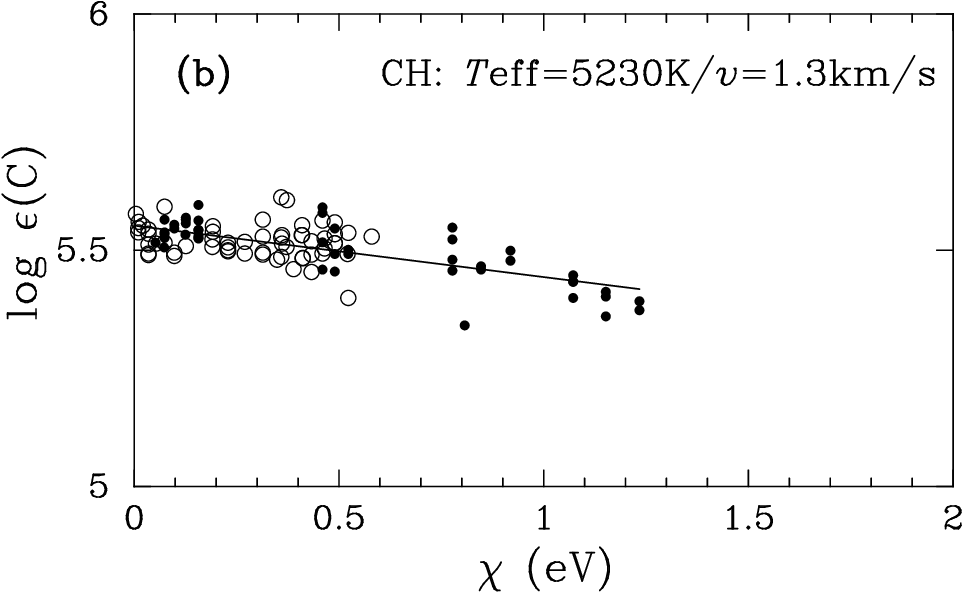} 
\plotone{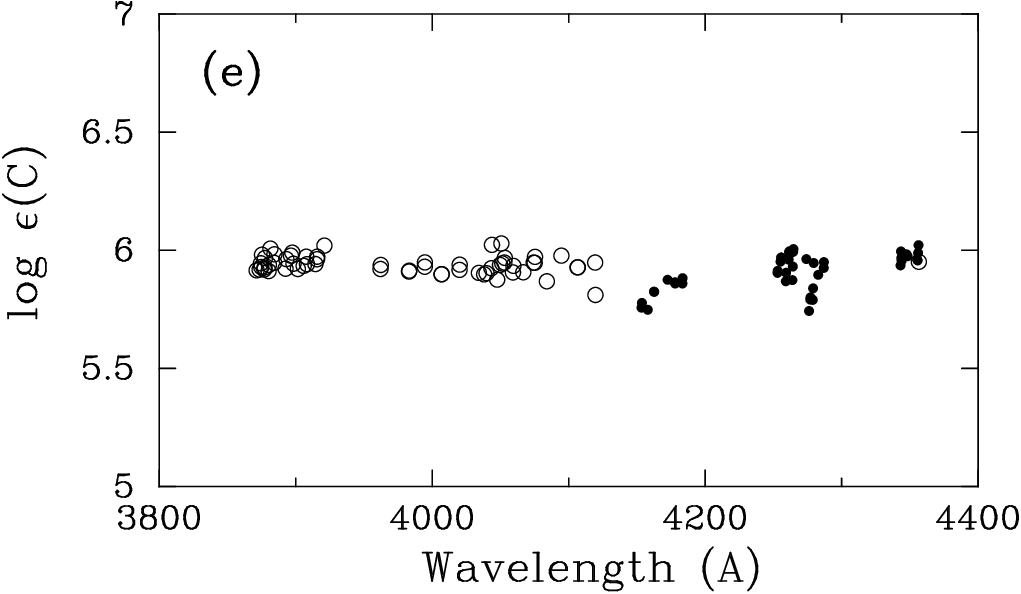} 
\plotone{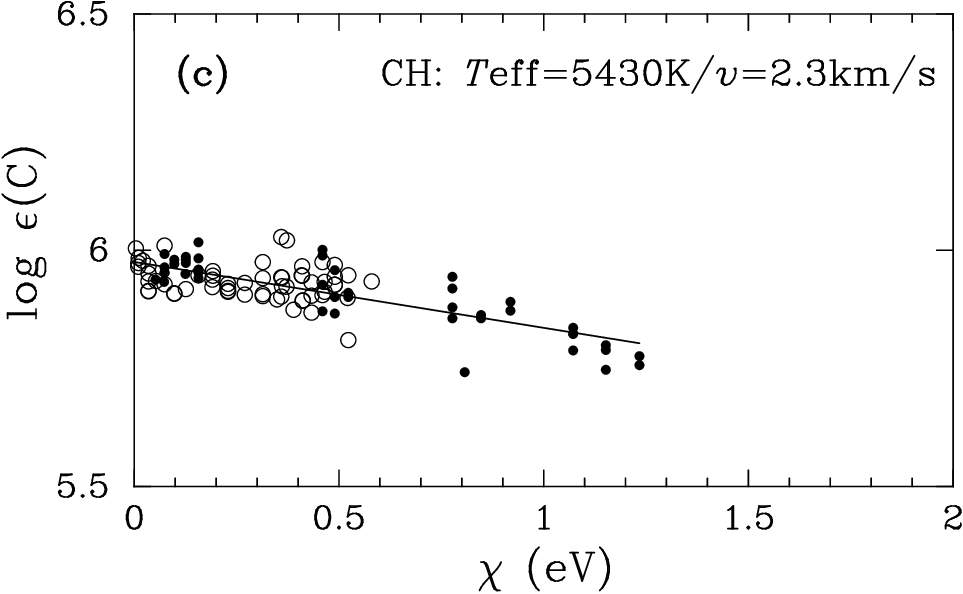} 
\plotone{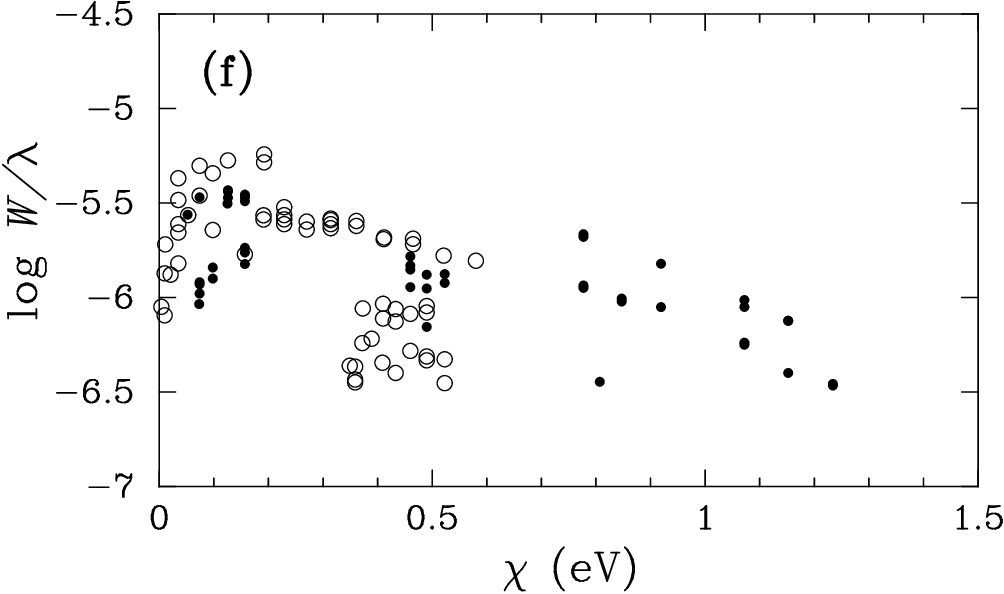} 
\caption{The same as Figure~\ref{fig:oh}, but for CH lines (filled
  circles: $A-X$ system; open circles: $B-X$ system). \label{fig:ch}}
\end{figure}

A similar analysis is made for CH lines as done for OH lines. The
results are shown in Figure~\ref{fig:ch}a. A clear correlation between the
carbon abundances derived from individual CH lines and their
excitation potentials is found as for OH lines. 
The null hypothesis that there is no correlation between the two values is
rejected by the regression analysis at the 99.9\% confidence level. The
slope $d\log\epsilon/d\chi$ is $-0.15 \pm 0.01$ dex/eV
(Fig. \ref{fig:ch}a).  The slope is slightly shallower,
$d\log\epsilon/d\chi = -0.14 \pm 0.01$ and $-0.11 \pm 0.01$ dex/eV if
larger micro-turbulent velocity ({\vt}$=2.3${\kms}) and lower
effective temperature ({\teff}$=5230$~K), respectively, are adopted
(Fig. 5b,c).

In the panels d and e of Figure~\ref{fig:ch}, no clear correlation
between the derived carbon abundances and wavelengths of CH lines, nor
line strengths, is found. This would be due to the fact that the
correlation between the wavelength and excitation potentials of CH
lines is not as clear as of OH lines (Fig. \ref{fig:ch}f).

\section{Discussion}

\subsection{OH and CH lines}

Our 1D-LTE abundance analysis for OH and CH molecular lines for the
extremely metal-poor star {\bd} demonstrates that the derived oxygen
and carbon abundances are dependent on the excitation potential with
slopes of $d\log\epsilon/d\chi =-0.15 \sim -0.2$ (dex/eV). This suggests that
the spectral lines with smaller excitation potentials are enhanced by
the existence of a cool layer in the atmosphere that is not reproduced
by the current 1D models. 3D atmosphere models predict that the upper
layer of photosphere is significantly cooler than that of the 1D
models. \citet{frebel08} report the correction of oxygen and carbon
abundances measured from OH and CH molecular lines using 3D model
atmospheres as a function of excitation potentials. They estimate that
the abundances derived from 3D models are lower than those from 1D
models by $\sim 0.75$~dex and by $\sim 0.55$~dex for lines with $\chi
\sim 0.0$ and 1.0~dex, respectively. Namely, the 3D corrections
explain the slopes of $d\log\epsilon/d\chi \sim -0.2$~dex/eV found by
our 1D analysis.

\citet{dobrovolskas13} also investigated line formation for molecules
including OH and CH based on the 3D model atmospheres for red giants
with {\teff}$\sim$ 5000~K. Their models for [Fe/H]$=-3$ also predict
similar trends of 3D corrections, as a function of excitation
potentials, for these molecules that are similar to those of
\citet{frebel08}. More quantitatively, the correction for CH lines
with $\chi\sim 0$ shown in Figure 9 of \citet{dobrovolskas13} is
smaller than those of \citet{frebel08}. This difference might be
partially due to the difference of the effective temperature and
metallicity studied by the two works. The shallower slope found for CH
than for OH in our result supports the prediction by \citet{dobrovolskas13}.

\subsection{Fe lines}

A similar dependence of the derived abundances on excitation
potentials of spectral lines is also expected for Fe
lines. \citet{collet07} estimate that the correction reaches
$-1.0$~dex for low excitation lines of \ion{Fe}{1}, and the slope
$d\log\epsilon/d\chi$ could be about $-0.2$ dex/eV for very metal-poor
stars. Such a clear trend is, however, not found in our previous
measurements of Fe lines \citep[Figure 5 of ][]{ito13}. A weaker
trend, $d\log\epsilon/d\chi \sim -0.05$ dex/eV, is found in the
derived Fe abundances from \ion{Fe}{1} lines. This trend, however,
almost disappears if a slightly low {\teff} is adopted in the analysis
\citep{ito13}.

Figure 6 shows the Fe abundances derived from individual \ion{Fe}{1}
lines as functions of excitation potentials and equivalent
widths. Large scatter is found in Fe abundances derived from lines
with small excitation potentials shown by filled and open circles in
the figure. Abundances from these lines show some dependence on line
strengths: results from lines with equivalent widths larger than
80~m{\AA} exhibit lower abundances. Such trends suggest that spectral
line formation is not well described by 1D LTE models.

The recent study by \citet{dobrovolskas13} predicts smaller
corrections for Fe abundances derived from \ion{Fe}{1} lines than that
of \citet{collet07}: their corrections are smaller than 0.2~dex for
lines with $\chi \geq 2$~eV.  The correction for Fe abundances from
\ion{Fe}{1} lines with $\chi = 0$~eV is, however, predicted to be
still as large as 0.6~dex. The abundances derived from lines with
$\chi\sim 0$ eV for {\bd} show scatter as large as 0.4~dex
\citep{ito13}. The large scatter of abundances from low excitation
\ion{Fe}{1} lines might indicate that these lines are particularly
affected by the 3D effect. These lines could also be affected by NLTE
effects. \citet{bergemann12} conducted NLTE analyses for Fe lines of
metal-poor stars using the spatial and temporal average stratification
from 3D hydrodynamic models, as well as 1D static model
atmospheres. They demonstrated for the cases of metal-poor stars like
the subgiant HD~140283 that the analysis with the mean 3D models
obtains lower Fe abundances on average, with smaller scatter of Fe
abundances from individual \ion{Fe}{1} lines, than those obtained from
1D models. On the other hand, the NLTE analysis using the 3D models
derives higher Fe abundances than those obtained by a LTE analysis. If
similar effects are working in the subgiant with lower metallicity
like {\bd}, the averaged Fe abundances derived by the 1D LTE analysis
is not significantly corrected by the 3D NLTE analysis, whereas the
scatter found in abundances from individual lines with small
excitation potentials becomes smaller.

We note that the Fe abundance obtained from \ion{Fe}{2} lines by our
analysis agrees well with that from \ion{Fe}{1} lines even if we adopt
the gravity ($\log g=3.3$) estimated adopting the parallax of this
star to be 4.88~mas that is independent of the spectroscopic analysis
of Fe lines \citep{ito13}. The NLTE and 3D effects estimated for
\ion{Fe}{2} lines are much smaller than those for neutral lines in
general \citep{collet07, bergemann12, dobrovolskas13}.  The agreement
of Fe abundances obtained from lines of the two ionization stages
using the 1D model suggests that the corrections by 3D NLTE analysis
for \ion{Fe}{1} lines is not as large as those estimated by 3D LTE models.

%

\subsection{Carbon and oxygen abundances of {\bd}}

The carbon and oxygen abundances of {\bd} derived from the lines with
relatively large excitation potentials ($\chi \sim 1.0$) 
are $\log \epsilon$(C)=5.8 and $\log
\epsilon$(O)=6.3, respectively (Fig. \ref{fig:ch} and
\ref{fig:oh}). Adopting the solar abundances $\log
\epsilon$(C)$_{\odot}$=8.43 and $\log \epsilon$(O)$_{\odot}$=8.69
\citep{asplund09} as well as [Fe/H]$=-3.8$ for {\bd}, the abundance
ratios are [C/Fe]$=+1.2$ and [O/Fe]$=+1.4$. The difference of oxygen
abundances obtained from OH lines with $\chi =1.0$ (eV) by 3D and 1D
model atmospheres ($\Delta_{\rm 3D-1D}$) is approximately $-0.4$~dex
\citep[Fig. 9 of ][]{dobrovolskas13}. Similarly, the $\Delta_{\rm
  3D-1D}$ of carbon abundances from CH lines with $\chi = 0.0$ and 1.0
(eV) are $-0.2$ -- $-0.1$~dex. Including these corrections, [C/Fe] and
      [O/Fe] of {\bd} are about 1.0~dex, indicating that carbon and
      oxygen are clearly enhanced in this object. We note that the C/O
      ratio obtained from CH and OH molecular lines estimated by 1D
      model atmospheres is robust.

The carbon abundance of {\bd} is measured using the \ion{C}{1} lines
at 1.068--1.069~$\mu$m by an NLTE analysis based on a 1D model atmosphere
by \citet{takeda13}, obtaining $\log \epsilon$(C)$=5.69$. This is
slightly lower than the carbon abundance obtained from CH lines with
1D models, but agrees well with the value corrected for the 3D effects
(see above). \citet{takeda13} mentioned, however, that the derived
carbon abundance from the \ion{C}{1} lines is sensitive to the
treatment of NLTE effects. Hence, we cannot conclude that the estimate
of the 3D effects for the carbon abundances from CH lines are
quantitatively correct from the comparison with the carbon abundance
from the \ion{C}{1} lines. However, the agreement of the abundances
within 0.1--0.2~dex from CH and \ion{C}{1} lines is supportive for the
reliable estimates of carbon abundance of this object.

\section{Summary} 

{\bd} is an extremely metal-poor subgiant with excesses of carbon and
oxygen. Oxygen and carbon abundances derived from individual OH and CH
lines by our analysis based on 1D LTE model atmospheres are dependent on the
excitation potentials. This result might indicate existence of cool
layer in the atmosphere which is not described by 1D models. Indeed,
the dependence found for oxygen and carbon abundances on excitation
potentials, $d\log \epsilon/d\chi \sim -0.15 \sim -0.2$ dex/eV, is
explained by the corrections estimated from 3D models. 


While the behavior of Fe absorption lines is complicated, our analysis
of OH and CH molecular lines qualitatively supports the prediction of
the 3D models on the temperature structure of the surface
atmosphere. For more quantitative discussion, investigations of the
3D models focusing on the stellar parameters of {\bd} are desired.

\acknowledgments




{\it Facilities:} \facility{Subaru(HDS)}.

\acknowledgments

W.A. was supported by the Grants-in-Aid for Science
Research of JSPS (20244035).

\clearpage



\begin{figure}
\epsscale{.70}
\plotone{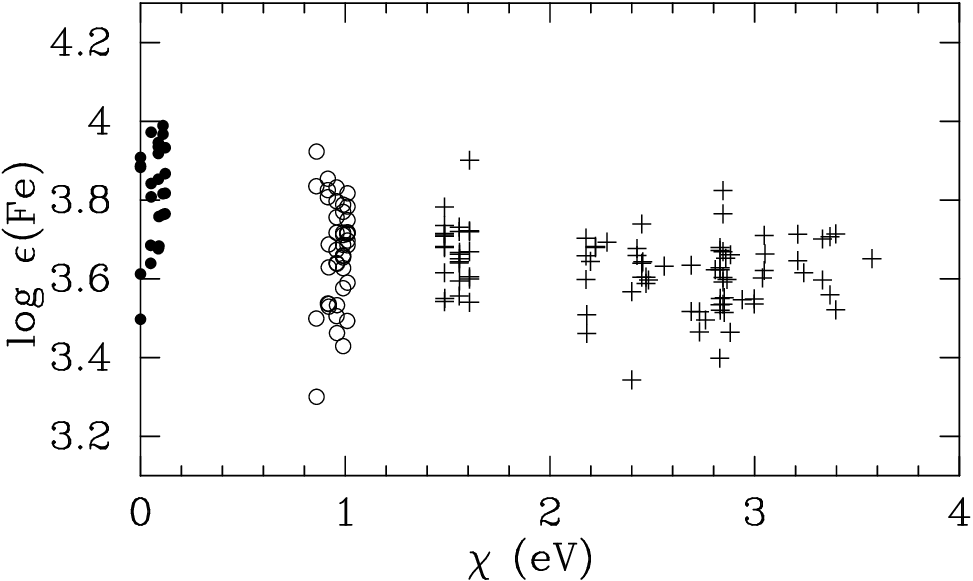}
\plotone{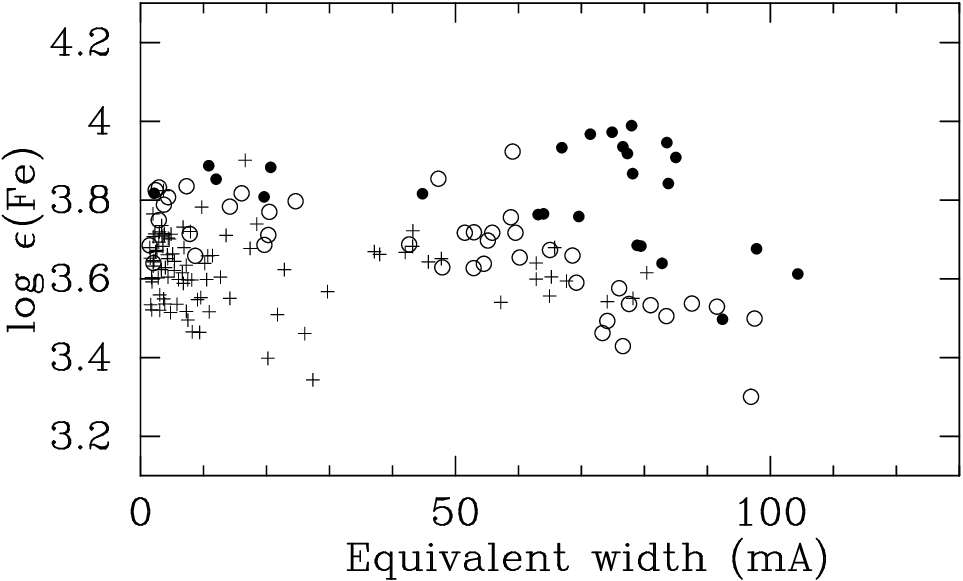}
\caption{Fe abundances derived from individual \ion{Fe}{1} lines as a
  function of excitation potentials (upper panel) and equivalent
  widths (lower panel). Filled and open circles are results from lines
  with $\chi<0.5$ (eV) and $0.5<\chi<1.2$ (eV), respectively. In each
  $\chi$ ranges, the abundances derived from lines with equivalent
  widths larger than 80~m{\AA} show lower abundances than those from
  weaker lines, which yields large scatter in the abundances. See text
  for more details. \label{fig:fe1}}
\end{figure}










\begin{deluxetable}{cccc}
\tabletypesize{\scriptsize}
\tablewidth{0pt}
\tablecaption{Equivalent widths of OH $A-X$ lines measured for {\bd}  \label{tab:oh}}
\tablehead{
\colhead{Wavelength ({\AA})} & \colhead{$\chi$ (eV)} &  \colhead{$\log gf$} &  \colhead{$W$ (m{\AA})}
}
\startdata
\multicolumn{4}{c}{OH $A-X$ 0--0 system}  \\
\tableline
  3081.254 &      0.762 &     -1.882 &     21.1 \\ 
  3084.048 &      0.016 &     -3.162 &     12.6 \\ 
  3084.895 &      0.843 &     -1.874 &     18.8 \\ 
  3086.224 &      0.925 &     -1.852 &     14.4 \\ 
  3086.391 &      0.010 &     -2.482 &     37.0 \\ 
  3087.342 &      0.095 &     -1.896 &     58.9 \\ 
  3089.008 &      0.928 &     -1.869 &     15.0 \\ 
  3090.862 &      1.013 &     -1.850 &     12.8 \\ 
  3092.397 &      0.164 &     -1.782 &     65.5 \\ 
  3093.724 &      0.016 &     -2.860 &     24.6 \\ 
  3094.619 &      0.134 &     -1.899 &     63.4 \\ 
  3095.344 &      0.205 &     -1.738 &     64.9 \\ 
  3095.545 &      0.205 &     -3.121 &      7.6 \\ 
  3096.650 &      0.170 &     -3.098 &     11.4 \\ 
  3098.589 &      0.250 &     -1.700 &     63.3 \\ 
  3099.414 &      0.210 &     -1.788 &     65.7 \\ 
  3101.230 &      0.067 &     -2.205 &     48.0 \\ 
  3102.144 &      0.300 &     -1.667 &     66.5 \\ 
  3104.351 &      1.202 &     -1.868 &     10.4 \\ 
  3105.666 &      0.304 &     -1.708 &     47.6 \\ 
  3106.019 &      0.354 &     -1.639 &     65.4 \\ 
  3106.546 &      0.095 &     -2.142 &     49.2 \\ 
  3107.855 &      1.297 &     -1.858 &      6.0 \\ 
  3110.226 &      0.412 &     -1.615 &     56.8 \\ 
  3110.530 &      1.300 &     -1.873 &      9.9 \\ 
  3113.365 &      0.415 &     -1.649 &     55.3 \\ 
  3114.773 &      0.474 &     -1.594 &     54.5 \\ 
  3123.948 &      0.205 &     -2.003 &     58.0 \\ 
  3127.687 &      0.612 &     -1.589 &     51.7 \\ 
  3128.286 &      0.210 &     -2.074 &     56.6 \\ 
  3130.281 &      0.250 &     -1.969 &     65.4 \\ 
  3130.570 &      0.683 &     -1.549 &     52.7 \\ 
  3133.228 &      0.686 &     -1.575 &     44.1 \\ 
  3134.344 &      0.255 &     -2.031 &     54.0 \\ 
  3136.895 &      0.299 &     -1.939 &     51.0 \\ 
  3139.169 &      0.763 &     -1.564 &     34.0 \\ 
  3140.506 &      0.304 &     -3.041 &      6.7 \\ 
  3140.734 &      0.304 &     -1.994 &     46.0 \\ 
  3145.520 &      0.845 &     -1.555 &     31.5 \\ 
  3151.003 &      0.411 &     -1.890 &     42.0 \\ 
  3159.505 &      1.017 &     -1.545 &     21.7 \\ 
  3166.337 &      0.538 &     -1.853 &     35.3 \\ 
  3169.613 &      0.542 &     -1.891 &     32.7 \\ 
  3172.992 &      1.202 &     -1.526 &     16.2 \\ 
  3174.483 &      0.608 &     -1.839 &     37.2 \\ 
  3175.301 &      1.204 &     -1.545 &     13.8 \\ 
  3181.646 &      1.300 &     -1.530 &     14.4 \\ 
  3182.962 &      0.682 &     -1.827 &     28.5 \\ 
  3183.921 &      1.302 &     -1.549 &     13.6 \\ 
  3194.847 &      0.762 &     -1.848 &     23.7 \\ 
  3200.488 &      1.505 &     -1.546 &     20.0 \\ 
  3200.959 &      0.840 &     -1.810 &     28.0 \\ 
  3203.975 &      0.843 &     -1.839 &     20.4 \\ 
  3210.500 &      0.925 &     -1.805 &     49.8 \\ 
  3220.418 &      1.013 &     -1.802 &     14.5 \\ 
  3223.366 &      1.016 &     -1.828 &     11.5 \\ 
  3223.731 &      1.723 &     -1.588 &      3.6 \\ 
  3230.729 &      1.104 &     -1.802 &      9.9 \\ 
  3233.654 &      1.107 &     -1.826 &     10.9 \\ 
  3241.448 &      1.199 &     -1.803 &      8.1 \\ 
  3252.594 &      1.297 &     -1.807 &      5.8 \\ 
  3255.492 &      1.300 &     -1.829 &      5.2 \\ 
  3264.185 &      1.398 &     -1.813 &      4.0 \\ 
\tableline
\multicolumn{4}{c}{OH $A-X$ 1--1 system} \\
\tableline
  3122.962 &      0.507 &     -2.555 &     10.3 \\ 
  3127.042 &      0.571 &     -2.442 &     10.5 \\ 
  3127.353 &      0.735 &     -2.269 &     10.8 \\ 
  3128.060 &      0.541 &     -2.511 &     21.2 \\ 
  3128.518 &      0.786 &     -2.242 &     16.3 \\ 
  3129.538 &      0.516 &     -2.597 &      9.3 \\ 
  3136.178 &      0.452 &     -2.570 &     17.9 \\ 
  3139.794 &      0.485 &     -2.279 &     20.4 \\ 
  3141.911 &      0.507 &     -2.186 &     22.0 \\ 
  3146.947 &      0.565 &     -2.052 &     24.5 \\ 
  3148.430 &      0.516 &     -2.277 &     17.6 \\ 
  3153.204 &      0.640 &     -1.960 &     36.9 \\ 
  3163.215 &      0.534 &     -2.351 &     14.8 \\ 
  3165.150 &      0.783 &     -1.871 &     21.5 \\ 
  3168.983 &      0.565 &     -2.298 &     16.1 \\ 
  3169.174 &      0.541 &     -2.472 &      9.3 \\ 
  3169.866 &      0.838 &     -1.851 &     17.6 \\ 
  3173.198 &      0.842 &     -1.885 &     18.0 \\ 
  3180.472 &      0.961 &     -1.822 &     15.2 \\ 
  3183.510 &      0.965 &     -1.851 &     20.1 \\ 
  3186.389 &      1.028 &     -1.812 &     13.7 \\ 
  3188.062 &      0.683 &     -2.182 &     12.1 \\ 
  3189.311 &      1.032 &     -1.840 &      9.7 \\ 
  3195.555 &      1.102 &     -1.832 &     12.6 \\ 
  3199.108 &      0.735 &     -2.209 &     12.9 \\ 
  3199.523 &      1.174 &     -1.803 &     22.7 \\ 
  3206.235 &      0.786 &     -2.180 &     10.6 \\ 
  3206.771 &      1.252 &     -1.804 &      7.2 \\ 
  3210.042 &      0.837 &     -2.110 &     10.8 \\ 
  3213.738 &      0.842 &     -2.155 &     11.7 \\ 
  3218.061 &      0.897 &     -2.093 &     10.1 \\ 
  3229.891 &      0.964 &     -2.118 &      8.9 \\ 
  3238.552 &      1.031 &     -2.104 &      6.6 \\ 
  3240.750 &      1.597 &     -1.836 &      3.8 \\ 
  3253.875 &      1.172 &     -2.057 &      6.1 \\ 
  3263.826 &      1.250 &     -2.055 &      3.2 \\ 
\tableline
\multicolumn{4}{c}{OH $A-X$ 2--2 system} \\
\tableline
  3206.518 &      0.953 &     -2.387 &      4.2 \\ 
  3221.127 &      1.059 &     -2.295 &      3.9 \\ 
\enddata
\end{deluxetable}

\begin{deluxetable}{cccc}
\tabletypesize{\scriptsize}
\tablewidth{0pt}
\tablecaption{Equivalent widths of CH $B-X$ and $A-X$ lines measured for {\bd}  \label{tab:ch}}
\tablehead{
\colhead{Wavelength ({\AA})} & \colhead{$\chi$ (eV)} &  \colhead{$\log gf$} &  \colhead{$W$ (m{\AA})}
}
\startdata
\multicolumn{4}{c}{CH $B-X$ system}  \\
\tableline
  3871.363 &      0.098 &      -1.68 &      8.8 \\ 
  3873.557 &      0.035 &      -1.94 &      5.9 \\ 
  3874.543 &      0.191 &      -1.55 &     10.0 \\ 
  3874.782 &      0.191 &      -1.51 &     10.5 \\ 
  3875.311 &      0.021 &      -2.08 &      5.1 \\ 
  3876.980 &      0.229 &      -1.51 &      9.5 \\ 
  3877.193 &      0.229 &      -1.49 &     10.1 \\ 
  3877.467 &      0.010 &      -2.30 &      3.1 \\ 
  3880.186 &      0.270 &      -1.49 &      8.9 \\ 
  3880.387 &      0.270 &      -1.47 &      9.8 \\ 
  3881.478 &      0.004 &      -2.30 &      3.5 \\ 
  3884.225 &      0.314 &      -1.47 &      9.1 \\ 
  3884.414 &      0.314 &      -1.45 &     10.2 \\ 
  3892.585 &      0.035 &      -1.58 &     12.8 \\ 
  3893.064 &      0.035 &      -1.49 &     16.6 \\ 
  3896.530 &      0.010 &      -2.08 &      5.2 \\ 
  3897.480 &      0.011 &      -1.93 &      7.4 \\ 
  3898.389 &      0.074 &      -1.35 &     19.4 \\ 
  3901.591 &      0.098 &      -1.35 &     17.7 \\ 
  3905.902 &      0.126 &      -1.25 &     20.8 \\ 
  3907.766 &      0.035 &      -1.82 &      8.6 \\ 
  3908.268 &      0.035 &      -1.74 &      9.6 \\ 
  3914.419 &      0.053 &      -1.67 &     10.7 \\ 
  3915.609 &      0.192 &      -1.22 &     20.3 \\ 
  3915.790 &      0.192 &      -1.18 &     22.4 \\ 
  3921.072 &      0.074 &      -1.61 &     13.6 \\ 
  3962.173 &      0.229 &      -1.46 &     10.8 \\ 
  3962.389 &      0.229 &      -1.43 &     11.9 \\ 
  3983.001 &      0.314 &      -1.41 &      9.7 \\ 
  3983.191 &      0.314 &      -1.39 &     10.2 \\ 
  3994.501 &      0.361 &      -1.39 &      9.6 \\ 
  3994.691 &      0.361 &      -1.38 &     10.1 \\ 
  4006.805 &      0.411 &      -1.38 &      8.2 \\ 
  4006.994 &      0.412 &      -1.37 &      8.3 \\ 
  4020.020 &      0.465 &      -1.37 &      7.7 \\ 
  4020.186 &      0.465 &      -1.36 &      8.2 \\ 
  4034.234 &      0.521 &      -1.36 &      6.7 \\ 
  4038.113 &      0.349 &      -2.14 &      1.8 \\ 
  4039.860 &      0.359 &      -2.14 &      1.7 \\ 
  4043.332 &      0.372 &      -2.02 &      2.3 \\ 
  4043.825 &      0.373 &      -1.93 &      3.5 \\ 
  4047.667 &      0.389 &      -1.93 &      2.5 \\ 
  4049.555 &      0.580 &      -1.36 &      0.0 \\ 
  4049.715 &      0.580 &      -1.36 &      6.3 \\ 
  4050.685 &      0.359 &      -2.35 &      1.4 \\ 
  4051.358 &      0.359 &      -2.25 &      1.5 \\ 
  4052.936 &      0.410 &      -1.87 &      3.1 \\ 
  4053.263 &      0.410 &      -1.81 &      3.8 \\ 
  4059.206 &      0.433 &      -1.82 &      3.0 \\ 
  4059.484 &      0.433 &      -1.78 &      3.5 \\ 
  4066.800 &      0.460 &      -1.75 &      3.3 \\ 
  4074.320 &      0.409 &      -2.11 &      1.8 \\ 
  4075.112 &      0.490 &      -1.75 &      3.4 \\ 
  4075.320 &      0.490 &      -1.74 &      3.7 \\ 
  4084.001 &      0.433 &      -2.06 &      1.6 \\ 
  4094.682 &      0.460 &      -2.02 &      2.1 \\ 
  4106.456 &      0.490 &      -1.99 &      1.9 \\ 
  4106.713 &      0.490 &      -1.97 &      2.0 \\ 
  4119.432 &      0.523 &      -1.97 &      1.9 \\ 
  4119.669 &      0.523 &      -1.96 &      1.5 \\ 
\tableline
\multicolumn{4}{c}{CH $A-X$ system} \\
\tableline
  4153.408 &      1.234 &      -1.16 &      1.4 \\ 
  4153.634 &      1.234 &      -1.17 &      1.5 \\ 
  4158.012 &      1.152 &      -1.17 &      1.7 \\ 
  4162.464 &      1.072 &      -1.17 &      2.4 \\ 
  4162.667 &      1.072 &      -1.18 &      2.3 \\ 
  4172.513 &      0.919 &      -1.19 &      3.7 \\ 
  4177.834 &      0.847 &      -1.21 &      4.1 \\ 
  4177.997 &      0.847 &      -1.22 &      4.0 \\ 
  4183.330 &      0.777 &      -1.22 &      4.7 \\ 
  4183.472 &      0.777 &      -1.23 &      4.8 \\ 
  4253.000 &      0.523 &      -1.51 &      5.1 \\ 
  4253.206 &      0.523 &      -1.47 &      5.7 \\ 
  4255.248 &      0.157 &      -1.46 &     14.4 \\ 
  4255.628 &      0.157 &      -1.50 &     13.8 \\ 
  4255.823 &      0.157 &      -1.46 &     14.9 \\ 
  4259.087 &      0.490 &      -1.54 &      4.8 \\ 
  4259.282 &      0.490 &      -1.50 &      5.6 \\ 
  4261.218 &      0.125 &      -1.54 &     13.4 \\ 
  4261.521 &      0.126 &      -1.49 &     15.8 \\ 
  4261.731 &      0.126 &      -1.54 &     14.4 \\ 
  4261.982 &      0.126 &      -1.49 &     15.5 \\ 
  4263.969 &      0.460 &      -1.57 &      4.8 \\ 
  4264.265 &      0.460 &      -1.53 &      6.0 \\ 
  4264.466 &      0.460 &      -1.57 &      6.3 \\ 
  4264.711 &      0.460 &      -1.53 &      7.1 \\ 
  4274.186 &      0.074 &      -1.56 &     14.5 \\ 
  4276.083 &      0.807 &      -1.58 &      1.5 \\ 
  4277.010 &      1.152 &      -0.93 &      3.2 \\ 
  4277.230 &      1.152 &      -0.94 &      3.2 \\ 
  4278.855 &      1.072 &      -0.94 &      3.8 \\ 
  4279.055 &      1.072 &      -0.95 &      4.2 \\ 
  4279.480 &      0.052 &      -1.67 &     11.7 \\ 
  4282.787 &      0.919 &      -0.97 &      6.5 \\ 
  4286.886 &      0.777 &      -1.00 &      9.0 \\ 
  4287.036 &      0.777 &      -1.01 &      9.3 \\ 
  4343.173 &      0.073 &      -2.14 &      4.0 \\ 
  4343.489 &      0.074 &      -2.14 &      4.6 \\ 
  4343.687 &      0.074 &      -2.05 &      5.2 \\ 
  4343.963 &      0.074 &      -2.05 &      5.1 \\ 
  4347.538 &      0.098 &      -2.02 &      5.5 \\ 
  4348.336 &      0.098 &      -1.95 &      6.3 \\ 
  4355.341 &      0.490 &      -1.84 &      3.0 \\ 
  4355.693 &      0.157 &      -1.85 &      6.5 \\ 
  4355.992 &      0.157 &      -1.79 &      8.0 \\ 
  4356.355 &      0.157 &      -1.85 &      7.5 \\ 
  4356.594 &      0.157 &      -1.79 &      7.4 \\ 
\enddata
\end{deluxetable}

\end{document}